\documentclass[aps,longbibliography,
  twocolumn,amsmath,
  amssymb,nofootinbib, superscriptaddress, preprintnumbers]{revtex4-1}

\usepackage{graphics}      
\usepackage{graphicx}      
\usepackage{longtable}     
\usepackage{url}           
\usepackage{bm}            
\usepackage{xcolor}
\usepackage{mathrsfs}
\usepackage[colorlinks,allcolors=blue]{hyperref}
\usepackage{blindtext}
\usepackage{hyperref}
\usepackage{float}
\usepackage{bbold}
\usepackage{lipsum}
\usepackage[normalem]{ulem}
\usepackage{orcidlink}
\usepackage{academicons}
\usepackage{xcolor}

\newcommand{\orcid}[1]{\href{https://orcid.org/#1}{\textcolor[HTML]{A6CE39}{\aiOrcid}}}

\begin{document}
\preprint{MPP-2022-108}

\title{Non-Standard Neutrino Self-Interactions Can Cause   Neutrino \\ Flavor Equipartition Inside the Supernova Core}

\newcommand*{\MPP}{\textit{\small{Max-Planck-Institut f\"ur Physik (Werner-Heisenberg-Institut), F\"ohringer Ring 6, 80805 M\"unchen, Germany}}}
\author{Sajad Abbar \orcidlink{0000-0001-8276-997X}   \\  \MPP } 


\begin{abstract}
We show that non-standard  neutrino self-interactions can lead to  total flavor 
equipartition in a dense neutrino gas, such as those expected in 
 core-collapse supernovae.
 In this first investigation of this phenomenon in the multi-angle scenario, 
 we demonstrate that such a  flavor equipartition can occur on very short scales, 
 and therefore very deep  inside the newly formed proto-neutron star,
 with a possible significant impact on the physics of core-collapse supernovae. 
 Our findings imply that    
 future galactic core-collapse supernovae can appreciably probe non-standard neutrino
  self-interactions, for certain cases even when they are many orders of magnitude smaller than the Standard Model terms.
  
 \end{abstract}

\maketitle

\section{Introduction}
Core-collapse supernova (CCSN) explosions, caused by the death of massive stars,
are among the most
energetic astrophysical settings~\cite{Colgate:1966ax, Bethe:1984ux, 
 Janka:2012wk, Burrows:2012ew}.
During the explosion, a huge amount of energy is reassessed of which  almost
99\% is in the form of neutrinos of all flavors. 
 Given the short duration of the burst,  the number density 
 of the neutrinos  in such environments is so huge that the neutrino-neutrino
 interactions can play a crucial role in their flavor evolution~\cite{Qian:1996xt, Pastor:2002we,duan:2006an, duan:2006jv, duan:2010bg,
  Chakraborty:2016yeg}.

 In this paper, 
 we study the impact of non-standard neutrino self-interactions ($\nu$NSSI)
 on their flavor evolution in dense neutrino media. 
 Such $\nu$NSSI  are allowed  in some of the beyond Standard Model (BSM) theories of particle physics~\cite{Bialynicka-Birula:1964ddi,Bardin:1970wq}.
 While a BSM scaler mediator results in the presence of a trivial mass term of the form $\bar\nu_\xi  \nu_\eta$ in the Lagrangian~\cite{Ge:2018uhz},
  a vector mediator   leads to the 
 effective Lagrangian $\mathcal{L}_{\mathrm{eff}} \supset G_{\mathrm{F}} [\mathsf{G}^{\alpha\beta} \bar\nu_\alpha \gamma^{\mu} \nu_\beta] [\mathsf{G}^{\xi\eta} \bar\nu_\xi \gamma_{\mu} \nu_\eta]$, which
 is very similar to the neutrino-neutrino interaction Lagrangian  of the SM except that it now allows for new interaction
  terms coupling neutrinos with different flavors via $\mathsf{G}^{\alpha\beta}$'s (see Ref.~\cite{Yang:2018yvk} for a discussion on how
 the effective single-particle Hamiltonian obtained from the Lagrangian can have a different structure). Needless to say, the SM can be 
 recovered by setting $\mathsf{G}^{\alpha\beta} = \delta_{\alpha\beta} $.
 The  non-standard  components of $\mathsf{G}$ are related to the vector mediator mass $m_V$ and the coupling strength $g_V$, by $|\mathsf{G}^{\alpha\beta}| \propto g_V^2/m_V^2$. 
  
  The current constraints on $\nu$NSSI are rather weak and mostly model dependent, namely  
  $|\mathsf{G}^{\alpha\beta}|$ can be in general a couple of orders of magnitude larger than one (see, e.g., Fig.~1 of Ref.~\cite{Berryman:2022hds}). 
  Though  laboratory constraints are expectedly looser, 
   stronger limits are feasible from the astrophysical neutrinos and the early Universe.
   
   As for the constraints from laboratory processes, 
   despite the fact that the number density of neutrinos is very low to allow for any direct detection of $\nu$NSSI,
   it can be indirectly probed by the measurements of other particles decay/interactions.
   Of particular interest is the measurement of the   invisible width of the
Z-boson~\cite{Bilenky:1992xn, Bilenky:1994ma}(see also a nice improvement of this in Ref.~\cite{Brdar:2020nbj}). 

$\nu$NSSI can also impact the physics of the early universe, in a number of ways.
For example, one can derive constraints on $\nu$NSSI noting that if the  mediator is light enough,
then it could be  in thermal equilibrium in the early universe, which  changes the number of relativistic degrees of freedom and
 can be probed   through Big Bang Nucleosynthesis~\cite{Ahlgren:2013wba}.
$\nu$NSSI can also affect Cosmic Neutrino Background properties, meaning that it can  be constrained 
via the Cosmic Microwave Background through its temperature angular power spectra~\cite{Archidiacono:2013dua, Cyr-Racine:2013jua}.

 $\nu$NSSI can be  constrained by the observation of  high-energy cosmic neutrinos as well.
 This comes from the fact  that  high-energy  neutrinos can be attenuated by their  propagation in the Cosmic Neutrino Background 
  if $\nu$NSSI is strong
   enough~\cite{Esteban:2021tub, Ioka:2014kca,Ng:2014pca,Ibe:2014pja,2014arXiv1408.3799B,DiFranzo:2015qea,Shoemaker:2015qul}.
 Besides,  CCSNe have also been considered as a probe of $\nu$NSSI~\cite{Kolb:1987qy, Shalgar:2019rqe, Chang:2022aas}.

 In order to study the flavor evolution of neutrinos in the presence of $\nu$NSSI, we solve the Liouville-von Neumann equation ($c=\hbar=1$)
\cite{Sigl:1992fn}
\begin{equation}
i d_t
\varrho_{\mathbf{p}} = \left[
  \frac{\mathsf{U}\mathsf{M}^2 \mathsf{U}^{\dagger}}{2E_\nu} + \mathsf{H}_{\mathrm{m}} +
  \mathsf{H}_{\nu \nu, \mathbf{p}} ,
  \varrho_{\mathbf{p}}\right],
\label{Eq:EOM}
\end{equation} 
where $\mathbf{p}$ is the neutrino momentum, $E_\nu=|\mathbf{p}|$, $\mathbf{v} = \mathbf{p}/E_\nu$, and $\mathsf{M}^2$ are the energy, velocity, and mass-square matrix of the neutrino, respectively, and $\mathsf{U}$ is the Pontecorvo–Maki–Nakagawa–Sakata matrix.
Moreover, $\mathsf{H}_{\mathrm{m}}$ is the contribution from the matter term which is proportional to matter (electron) density~\cite{Wolfenstein:1977ue,Mikheev:1986gs},
and $\mathsf{H}_{\nu \nu}$ is the neutrino potential stemming from the neutrino-neutrino forward 
scattering in the presence of $\nu$NSSI~\cite{sigl1993general,Blennow:2008er, Fuller:1987aa,Pantaleone:1992xh,Notzold:1988kx},
\begin{align}\label{eq:G}
\mathsf{H}_{\nu \nu, \mathbf{p}} = \sqrt2 G_{\mathrm{F}}
\int\!  \frac{\mathrm{d}^3p'}{(2\pi)^3} &( 1- \mathbf{v} \cdot \mathbf{v}')
\{\mathsf{G}(\varrho_{\mathbf{p}'} - \bar\varrho_{\mathbf{p}'})\mathsf{G}\nonumber \\
&+\mathsf{G}\ \mathrm{Tr}[(\varrho_{\mathbf{p}'} - \bar\varrho_{\mathbf{p}'})\mathsf{G}]   \},
\end{align}
where $\mathsf{G}=\mathsf{1}$, in the SM. The diagonal components of  $\mathsf{G}$ constitute the  flavor-preserving $\nu$NSSI, whereas the off-diagonal components show flavor-violating  $\nu$NSSI.

  The impact of $\nu$NSSI on the coherent scatterings of neutrinos in a dense neutrino gas
was first addressed in Ref.~\cite{Blennow:2008er}, in a single-angle two-flavor scenario. 
It was then investigated more in Refs.~\cite{Das:2017iuj, Dighe:2017sur}  
where a similar single-angle model was employed and it was shown that $\nu$NSSI can lead to flavor instabilities in both 
normal and inverted mass orderings. It was also particularly demonstrated that the flavor-violating  $\nu$NSSI lead to spectral splits
during the SN neutronization burst. 

In this paper, we provide the first multi-angle investigation of  $\nu$NSSI in a dense neutrino gas.
Our results  suggest that some of the most important insights obtained in the single-angle scenarios (Sec.~\ref{sec:sa}) should
be  artefacts of the limitations of such models. To be more specific, we show that the flavor-violating  $\nu$NSSI 
 can lead to flavor equipartition in a multi-angle neutrino gas on scales determined by the neutrino number density (Sec.~\ref{sec:ma}).

 \section{Two-flavor scenario}\label{sec:main}
 To study the flavor evolution of  neutrinos in the presence of $\nu$NSSI,
we consider a   neutrino gas in the  two-flavor 
scenarios ($\nu_e$ and $\nu_x$ with $x=\mu,\tau$, and their antiparticles) in which neutrinos 
are emitted uniformly with emission angles in the range  
$[-\vartheta_{\rm{max}}, \vartheta_{\rm{max}}]$. 
We then consider a number of examples in a single-energy single-angle one-dimensional (1D) gas, a single-energy multi-angle 1D gas,
a single-energy multi-angle 2D gas, and a multi-energy multi-angle 1D gas, respectively.
This sort of  models  have been
 extensively used in the literature (e.g., Ref.~\cite{Abbar:2018beu}). Also as it is usual in studying collective neutrino 
oscillations, we ignore the matter potential  assuming that it is constant and can be 
rotated away by transforming into a co-rotating frame. The dropping of the matter term
is specifically motivated once one considers neutrino flavor evolution deep inside the
PNS where on short scales, the neutrino gas and the ambient medium  can be considered to be
isotropic and homogenous  to a very good degree. In addition, except for our single-energy single-angle model,
we here assume that the strength of the week interaction defined as, $\mu= \sqrt2 G_{\mathrm{F}} n_{\nu_e}$, is constant,
and the vacuum frequency is set to be $|\omega|=1$~km$^{-1}$ for the single energy cases. 
 
 \begin{figure*} [tbh!]
\centering
\begin{center}
\includegraphics*[width=1.\textwidth, trim= 0 10 0 10, clip]{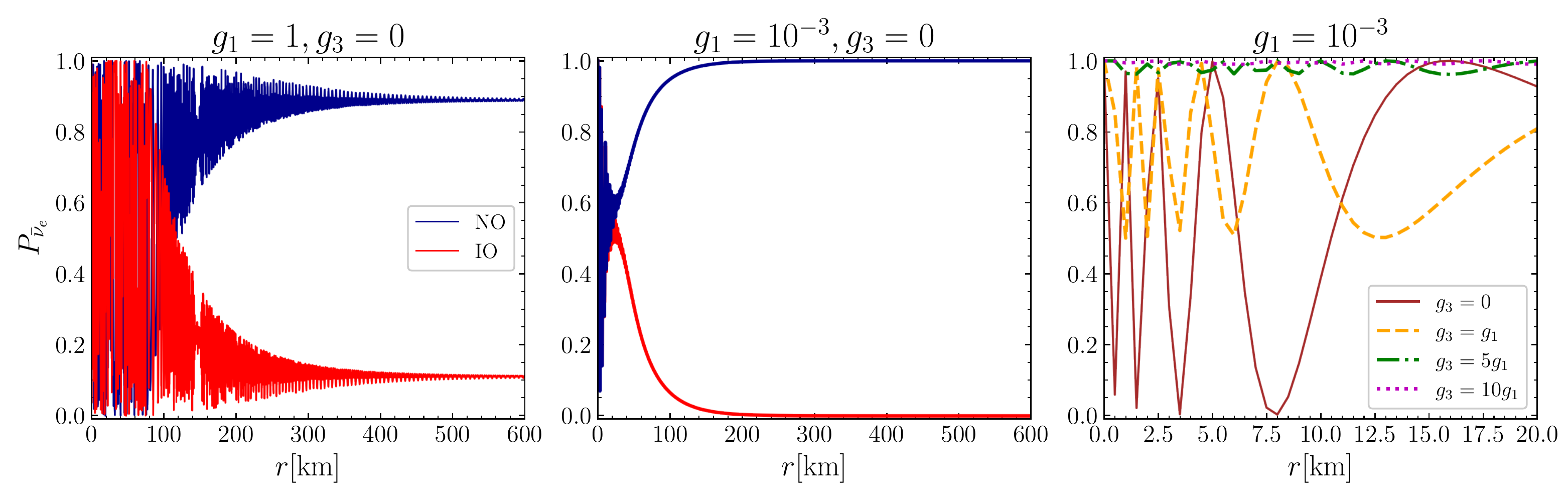}
\end{center}
\caption{
Survival probabilities of $\bar\nu_e$ as a function of the propagated distance, for
different values of $g_1$ and $g_3$.   The left and middle panels indicate  that in the  NO (IO) the flavor content of the neutrino
gas is almost completely survived (converted). The right panel on the other hand  shows the suppression caused by nonzero $g_3$.
Here we assume that $\mu = 3\times 10^4\ (10/r)^4$~km$^{-1}$ and
$n_{\bar\nu_e}/n_{\nu_e} = 0.8$. 
}
\label{fig:sa}
\end{figure*}

 In the two-flavor scenario, the  $\nu$NSSI matrix can be written as,
 \begin{equation}
\mathsf{G} = 
\left[ {\begin{array}{cc}
1+\gamma_{ee}  & \gamma_{ex}  \\
\gamma_{ex}^*  &  1+\gamma_{xx}    \\
\end{array} } \right].
\end{equation}
 Following Refs.~\cite{Blennow:2008er},  one can write the coupling matrix as, 
 \begin{equation}
 \mathsf{G} = \frac{1}{2}(g_0\mathsf{1} +\boldsymbol{g}\cdot\sigma),
 \end{equation}
where $g_0 = 2+\gamma_{ee}+\gamma_{xx}$, $g_1 = 2 \rm{Re}(\gamma_{ex})$, $g_2 = 2 \rm{Im}(\gamma_{ex}^*)$,
and $g_3 = \gamma_{ee} - \gamma_{xx}$, of which $g_2$ can be absorbed by a redefinition of the neutrino phases. 
Note that $g_0$ provides
a measure of the overall strength of the (nonstandard) weak interactions. From now on, we set $g_0=2$ so that
all the other quantities are normalised by $g_0$. 
The coupling matrix can now be written as, 
\begin{equation}
 \mathsf{G} = 
\left[ {\begin{array}{cc}
1+g_3  & g_1 \\
g_1  &  1- g_3    \\
\end{array} } \right].
\end{equation}
We here assume that $|g_3|<1$, so that the diagonal components of $\mathsf{G} $ are  positive.
This implies that 
one can always set $g_0$ in such a way that
the contributions from $\nu$NSSI to the diagonal components have   the same sign 
as those of the  SM, as might be naively expected because  the diagonal $\nu$NSSI 
contribution should be $\propto |g_V|^2$ and  the sign should not depend on the phase of $g_V$ (though we can not exclude the possibility of the opposite). 
This is unlike Ref.~\cite{Das:2017iuj}, where such a limitation was not considered 
and a number of noticeable phenomena were reported for the cases with $|g_3|>1$.

 \subsection{Single-angle scenario} \label{sec:sa}

For the sake of comparison and in order to build some useful insights, 
we here first provide our results of the single-angle calculations where we assume that in our model, all of the neutrino beams experience exactly the same
flavor evolution. To be consistent with the previous works~\cite{Blennow:2008er, Das:2017iuj}, we assume that 
the neutrino number density is decreasing as a function of the propagated distance such that 
$\mu = 3\times 10^4\ (10/r)^4$~km$^{-1}$, motivated by the supernova physics. As indicated in the left and middle
panels of  Fig.~\ref{fig:sa}, while for the normal mass ordering (NO) the final content of  $\bar\nu_e$ remains mostly unaffected
by the flavor-violating  $\nu$NSSI, it can be almost completely depleted in the inverted mass ordering (IO). These results 
show a fantastic agreement   with the results presented in Ref.~\cite{Blennow:2008er}, 
in spite of the difference between the geometries of the employed models.

The impact of nonzero flavor-preserving $\nu$NSSI is  indicated  in the right  panel of  Fig.~\ref{fig:sa}.
As can be clearly seen, the presence of nonzero $g_3$ tends to suppress the flavor conversions once $g_3\gtrsim g_1$.
As discussed in the following, this insight which was pointed out also in Ref.~\cite{Das:2017iuj}, survives in the multi-angle scenario.
 
 \begin{figure*} [tbh!]
\centering
\begin{center}
\includegraphics*[width=1.\textwidth, trim= 0 10 0 10, clip]{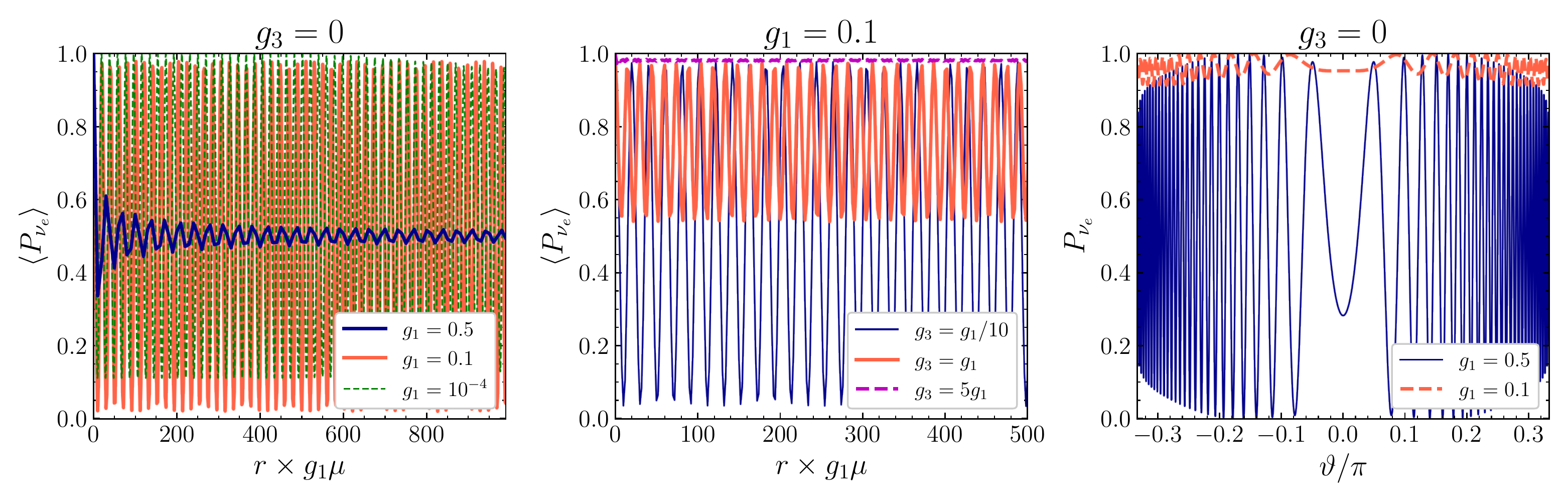}
\end{center}
\caption{
Left and middle panels: Angle-averaged survival probabilities of $\nu_e$ for different values of $g_1$ and $g_3$. 
Though high values of $g_3$ can suppress flavor conversions (middle), the neutrino gas reaches flavor equipartition
for large enough $g_1$ (left). While for larger $g_1$'s ($g_1\gtrsim 0.5$) an exact flavor equipartition can be reached, 
it occurs in an  average sense for smaller $g_1$'s. Right panel: Angular distributions of $\nu_e$ survival
probabilities. Note that $P_{\nu_e}$ is quite uneven  for $g_1= 0.5$.
Here we assume that $\mu = 10^6$~km$^{-1}$ and
$n_{\bar\nu_e}/n_{\nu_e} = 0.8$. Though here and in all other calculations in this study we set $n_{\nu_{x}} =0$,
we have confirmed that the equipartition discussed here is not affected once $n_{\nu_{x}} \neq 0$.
We have also confirmed that the results do not depend on the signs of  $g_1$ and $g_3$.
}
\label{fig:ma}
\end{figure*}

 \subsection{Multi-angle scenario}  \label{sec:ma}
 
 Having discussed the single-angle scenario, we now turn our focus to the multi-angle simulations.
 As indicated in the left panel of Fig.~\ref{fig:ma}, 
 even very tiny flavor-violating  $\nu$NSSI can lead to an almost perfect flavor equipartition 
 on scales $\sim (g_1\mu)^{-1}$, as long as $g_1\mu \gtrsim 100\ \omega_{\rm{atm}} $. 
 On the other hand, we have confirmed that 10\% flavor conversions can be induced once $g_1\mu \sim 10\ \omega_{\rm{atm}} $.

 Though the equipartition occurs in the exact manner for $g_1\gtrsim 0.5$,
 it holds on average for smaller $g_1$'s where the average is taken over a few oscillations.
 As a matter of fact, what matters here is the ratio $\mathsf{H}_{\nu \nu}^{\mathrm{off-diag}}/\mathsf{H}_{\nu \nu}^{\mathrm{diag}}$.
This can be  better understood from  the right panel of Fig.~\ref{fig:ma}, which presents the angular distributions of $\nu_e$ survival
probabilities for $g_1= 0.1$ and $0.5$, at $r\times g_1\mu =30$. While
  the angular distribution of $P_{\nu_e}$ shows large amplitude modulations for $g_1= 0.5$, 
 the different angle
beams are  oscillating in phase for $g_1= 0.1$. This explains 
the immediate flavor equipartition for $g_1= 0.5$. 

In addition,  nonzero flavor-preserving $\nu$NSSI can significantly 
suppress the flavor conversions provided that $g_3\gtrsim g_1$,
 as demonstrated  in the middle panel of Fig.~\ref{fig:ma}. Note that this is consistent with the insight obtained
in the single-angle scenario.

 \begin{figure} [tbh!]
\centering
\begin{center}
\includegraphics*[width=.52\textwidth, trim= 0 0 0 10, clip]{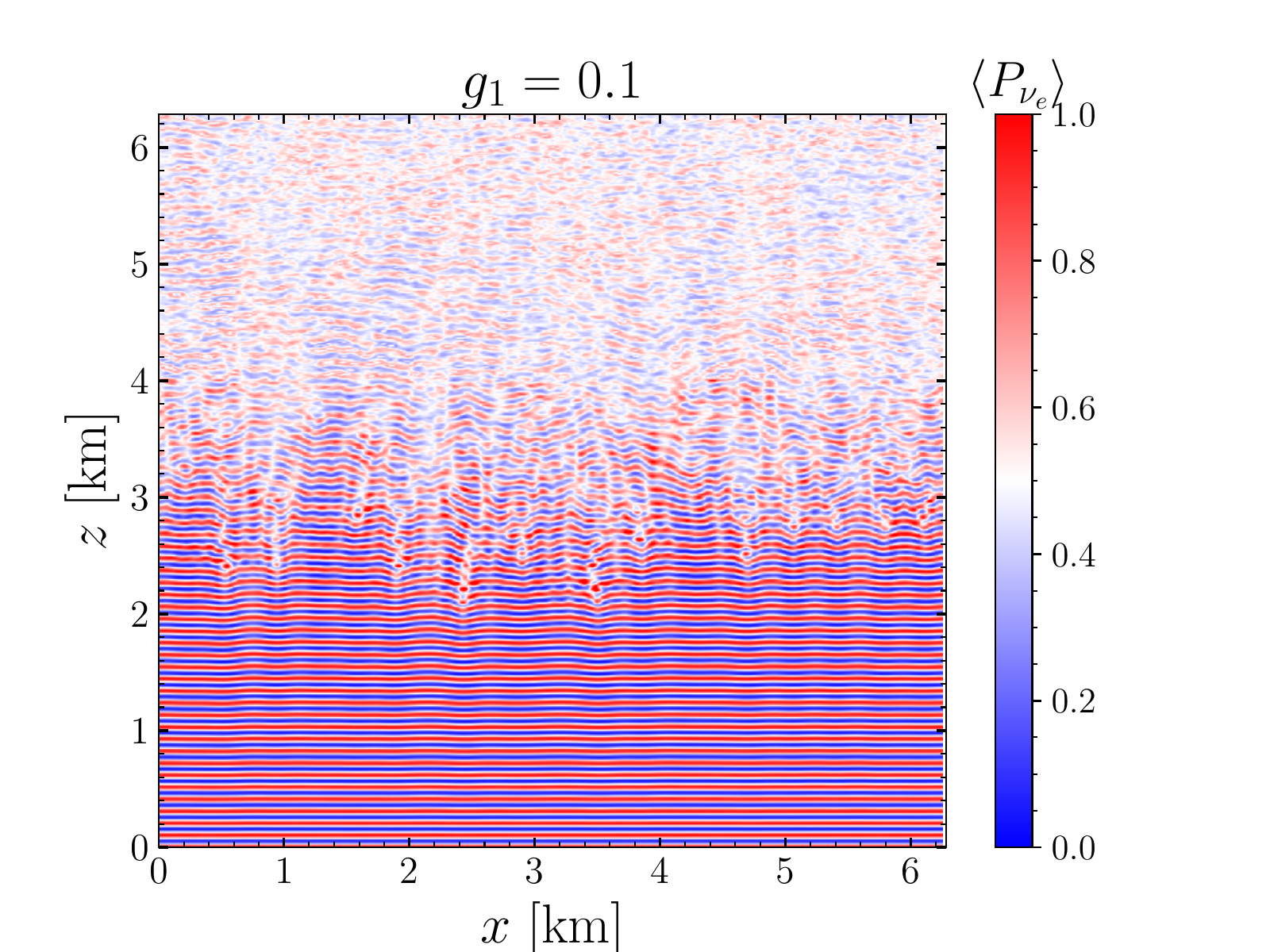}
\end{center}
\caption{
Angle-averaged $\nu_e$ survival probability in our 2D  single-energy multi-angle model. 
The  neutrino gas initially experiences a few  large amplitude oscillations  like what one
observes in 1D calculations. The coherence is then lost and an almost perfect flavor equipartition is reached.
Here we have assumed  $g_1=0.1$, $g_3=0$, $\mu=1000$~km$^{-1}$, and $n_{\bar\nu_e}=0$ (a pure $\nu_e$ gas). }
\label{fig:2D}
\end{figure}

Although for smaller values of the flavor-violating  $\nu$NSSI ($g_1\lesssim 0.5$) the flavor
equipartition occurs only on average in our 1D neutrino gas, the story could be 
different in MD models. This is specifically plausible considering the fact that in a MD neutrino
gas, each neutrino beam arriving at an arbitrary point (coming from different points on the source) 
should have already experienced a couple of large amplitude
oscillations. 
Assuming that these beams have evolved somewhat independently, one  should naively expect 
an exact equipartition  in a multi-angle MD neutrino gas even for smaller $g_1$'s.
This is illustrated clearly in Fig.~\ref{fig:2D},  for a calculation in which  $g_1=0.1$
and $g_3=0$. Here we have considered a 2D ($x$ and $z$) multi-angle neutrino gas in which a periodic boundary condition
is imposed along $x$, and the neutrino flavor evolution is followed along $z$. This model is similar to the one
used in Refs.~\cite{Abbar:2015mca, Martin:2019kgi, Abbar:2020ror}.
Though  
the neutrino gas experiences a few  large amplitude oscillations initially, the coherence is lost afterwards
and an almost perfect flavor equipartition is reached. 
 
 
 It is illuminating to note that the flavor equipartition discussed above (which is observed for 
 $g_1\gtrsim 0.5$ even in 1D models, but also observable for smaller $g_1$'s in MD models)  is purely a multi-angle effect. Indeed, it occurs  
 due to the occurence of large amplitude modulations in the angular distributions of neutrinos.

  Though the behaviour observed in Fig.~\ref{fig:2D}
is consistent with our insight that there should be an exact flavor equipartition in a MD gas, 
we alert the reader that we do not consider this topic as being settled 
since we could only
 run our 2D model simulations for very \emph{few} points in the parameter space of $\mu$-$\alpha$-$g_1$ ($\alpha = n_{\bar\nu_e}/n_{\nu_e}$). 
 This turns out to be indeed a very unstable 
 problem from the numerical point of view and we postpone a thorough investigation of this effect to a future work.

  \begin{figure} [tb!]
\centering
\begin{center}
\includegraphics*[width=.42\textwidth, trim= 0 10 0 10, clip]{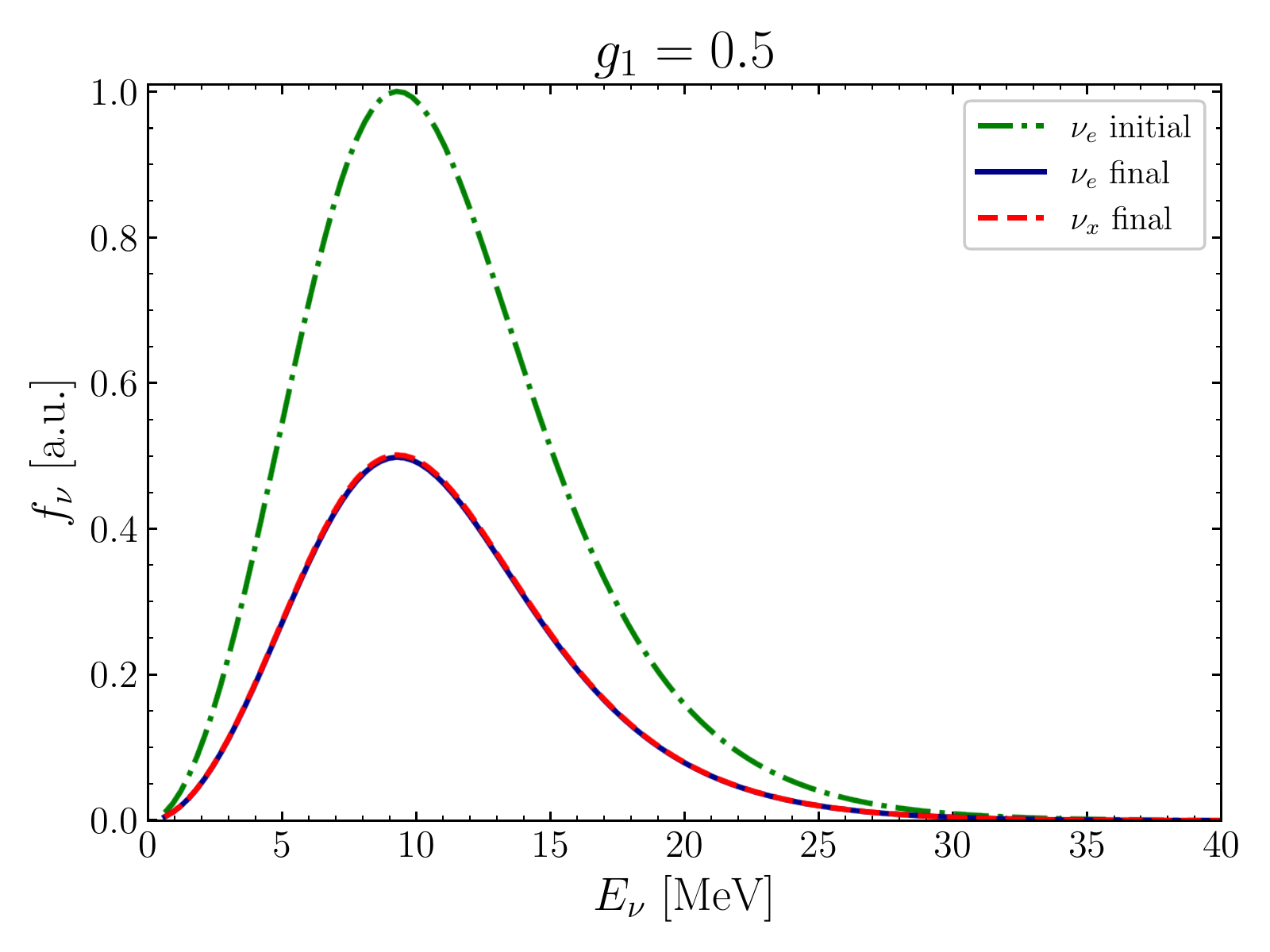}
\end{center}
\caption{
Initial and final neutrino spectra of a neutrino gas which  initially consists of a pure $\nu_e$ gas
(appropriate during the neutronization burst) in the presence of $\nu$NSSI.
As expected, the presence of flavor-violating  $\nu$NSSI leads to a total flavor equipartition during 
 the neutronization burst. Here we have assumed the initial neutrino gas to possess a Fermi-Dirac distribution
 equipped with a degeneracy parameter, though the choice of the initial spectra is arbitrary.
 Here we set  $\mu = 10^6$~km$^{-1}$. 
}
\label{fig:burst}
\end{figure}

 The  flavor equipartition discussed in this section is most relevant for the neutrino gas inside the PNS.
 This can be understood considering the following observations. First, the PNS
 is   the most sensitive zone to the $\nu$NSSI due to the highest neutrino number densities existing there. 
 Hence as long as $g_1\mu \gg \mathrm{max}(\omega,\ l_{\mathrm{col}}^{-1} )$, 
 the $\nu$NSSI-induced conversions
 can dominate the vacuum frequency oscillations and the collisional processes ($l_{\mathrm{col}}$ represents the collisional scales).
 In addition, if the neutrino gas is already in flavor equipartition at the neutrinosphere, then it should not change
 at larger radii (though we emphasise that this equipartition exists only in an average sense). 
 One should also note that for such a problem which shows variations on very short
 scales and in which MD effects are important, 1D large scales simulations, 
 as the one discussed in Fig.~\ref{fig:sa},
 should not be expeted to provide very useful insights.

The flavor equipartition  in the presence of $\nu$NSSI can even occur in the absence of
vacuum mixing, i.e., once $\omega, \theta_{\mathrm{V}}=0$, where  $ \theta_{\mathrm{V}}$ is the vacuum
mixing angle. This is because the  instability here is solely arising from the flavor-violating  $\nu$NSSI and
has nothing to do with the vacuum mixing. To be more specific, the occurrence of  flavor equipartition
is not sensitive to $\omega$ (or the neutrino mass ordering) as long as $g_1\mu \gg |\omega|$,

 One of the distinct results obtained in the single-angle scenario is the possibility of new 
 spectral splits during the SN neutronization burst due to the presence of flavor-violating  $\nu$NSSI~\cite{Das:2017iuj}  . 
 This is particularly interesting given the fact that one normally do not expect collective oscillations
 during this phase (unless the matter profile is too shallow, as discussed in Refs.~\cite{Duan:2007sh, Dasgupta:2008cd}). 
 However and as illustrated 
 in Fig.~\ref{fig:burst}, this behavior disappears in the multi-angle regime and instead, one has
  total flavor equipartition between $\nu_e$ and $\nu_{x}$, as expected from our discussions. 
  Here we have considered a multi-angle multi-energy 1D model which  initially consists
  of a pure $\nu_e$ gas. 
  Needless to say, such an equipartition should be
  easier to measure given the fact that it should exist for all the energy bins.
 
In Appendix~\ref{sec:linear}, we  provide a simple intuitive understanding of the   roles of $g_1$, $g_3$, and $\omega$ in flavor instabilities
 in the linear regime, which was previously missed
 in the literature.

\section{Conclusion}
We have studied neutrino flavor evolution in dense neutrino media in the presence of $\nu$NSSI, for the first time  in a multi-angle model.
We demonstrate that although some of the insights obtained in the single-angle scenario also apply to the multi-angle model,
some others should be an artefact of the single-angle approximation. In particular, we show that the dense neutrino gas can reach flavor equipartition on very short scales ( $\sim g_1^{-1} \mu^{-1}$)
in the presence $\nu$NSSI.
 We also
illustrate  that this effect can be suppressed by  flavor-preserving component if $g_3\gtrsim g_1$. 

The $\nu$NSSI-induced   flavor equipartition  is of most relevance to the neutrino flavor evolution inside the PNS and the accretion disks of neutron star merger (NSM) remnants.
This simply comes from the fact that the neutrino number density is very high in such environment. In addition, if neutrinos are already in flavor
equipartition on the surface of the neutrino emitter, then they might be expected to maintain this equipartition also at larger radii, unless
something destroys it.

The equipartition caused by the $\nu$NSSI inside the PNS can have important observational and theoretical consequences. 
On the one hand, considering specially the possible equipartition  during the neutronization burst, one can prob $g_1$ down to the values
$g_1 \gtrsim 10^{-6}$. On the other hand, considering the high neutrino number densities expected inside the 
PNS ($n_\nu \sim 10^{36}$~cm$^{-3}$), the  CCSN physics should be sensitive to $g_1 \gtrsim 10^{-8}$.

Though our study points out the interesting possibility of total flavor equipartition in a dense neutrino gas in the presence of $\nu$NSSI,
it has still several important limitations. 
First, though equipartition discussed in this work seems to be a robust phenomenon  in the two-flavor scenario,
the situation can be different in the three-flavor regime. This is specially expected considering the fact that in the three-flavor
scenario, there are more than one flavor-violating $\nu$NSSI parameters. Then the expected equilibrium  can in principle 
be different from equipartition. 
Second, the MD simulations of neutrino flavor evolution 
in the presence of  $\nu$NSSI, as we also pointed out in the text, are highly prone to numerical instabilities. Although we could run a few simulations 
and we observed  equipartition in MD models as presented in Fig.~\ref{fig:2D}, a systematic exploration of MD effects 
remains to be addressed in a future work.
In addition, neutrinos can also experience non-standard interactions with matter ($\nu$NSI). Though such interactions are well constrained in model-dependent 
scenarios~\cite{Proceedings:2019qno}, the model-independent bounds are relatively loose~\cite{Biggio:2009nt}. 
The impact of $\nu$NSI on neutrino flavor evolution has been extensively studied 
for the neutrino gas above the PNS and accretion disks of NSM remnants~\cite{Stapleford:2016jgz, Chatelain:2017yxx, Yang:2018yvk, Babu:2019iml, Fogli:2002xj}.
 Considering the equipartition induced by $\nu$NSSI, one might then wonder if strong $\nu$NSI can also lead to   
equipartition inside the PSN.


\section*{Acknowledgments}
I am grateful to Georg Raffelt and Meng-Ru Wu for very insightful discussions and their comments on the manuscript,
and Stefan Vogl, Manibrata Sen, and Irene Tamborra for valuable conversations. 
I acknowledges support by the Deutsche Forschungsgemeinschaft through
Sonderforschungbereich SFB 1258 Neutrinos and Dark Matter in Astro- and
Particle Physics (NDM).

\appendix

\section{Linear Stability Analysis}\label{sec:linear}
In this section, we show that some of the insights developed in the nonleianr regime in the previous section,
can be understood by linear stability analysis. For this purpose, we consider the simplest problem which can
 be solved analytically, i.e, a single-angle pure $\nu_e$  gas. 
 Note that although such a single-angle model cannot provide us with any insight on flavor equipartition,
 it still can be very useful in understanding the    roles of $g_1$, $g_3$, and $\omega$ in flavor instabilities.
 
 In the linear regime,
 where the flavor conversion is still insignificant ($|\delta|\ll1$), the neutrino density matrix can be written as,
 \begin{equation}
\rho= 
\left[ {\begin{array}{cc}
1  & \delta  \\
\delta^*  & 0    \\
\end{array} } \right],
\end{equation}
with the total Hamiltonian being
\begin{equation}
\mathsf{H} = 
\frac{\omega}{2}
\left[ {\begin{array}{cc}
1  & 0  \\
0  & -1    \\
\end{array} } \right]
+
\mu \big[   G \rho G+ G\ \rm{Tr}(\rho G)     \big],
\end{equation}
where  here $\mu$ captures all the information regarding the neutrino number density and the geometry.

Note that there is an important subtlety here: Unlike the conventional flavor stability analysis,
one \emph{cannot} remove the trace  of $\rho$ in the presence of $\nu$NSSI. This is obvious 
given the fact that in the presence of $\nu$NSSI, trace of $\rho$ leads to a  term $\propto G^2 + G\ \rm{Tr}(G)$
 in the Hamiltonian, which is not necessarily benign to the stability analysis. In the linear regime (ignoring terms nonlinear in $|\delta|$), the equation of motion
for $\delta$ then becomes,
 \begin{equation}\label{eq:linear}
i \partial_t \delta = (\omega + 4 \mu g_3^2 + 4 \mu g_3 - 2 \mu g_1^2) \delta -2\mu g_1^2 \delta^* - 2\mu g_1(1+g_3).
\end{equation}

 This equation is very interesting in several aspects. Firstly, there is a constant term which can be nonzero
 for nonzero $g_1$. This implies that for large enough $g_1$, $\delta$ can experience a sudden enhancement
 on scales $\propto \mu^{-1} g_1^{-1}(1+g_3)^{-1}$. However and though the right hand side of Eq.~(\ref{eq:linear}) is initially
 controlled by this term,  it can become finally subdominant  once $|\delta|$ grows enough (which might be already in the nonlinear regime).
 
 In addition, Eq.~(\ref{eq:linear}) couples $\delta$ to $\delta^*$, which is 
 different from the conventional linearised equation of motion for the flavor perturbations. This implies that the
 real and imaginary parts of $\delta$ can behave differently in the linear regime. 
 
 In order to get an impression of the stability of the neutrino gas in the presence of   $\nu$NSSI,
 we assume that $\delta$ is already in a regime where the $2\mu g_1(1+g_3)$ term can be ignored.
  Then one can find the following ordinary eigenvalue equation in terms of the  real and imaginary 
 components of $\delta$:
 \begin{equation}
 \partial_t
 \left[ {\begin{array}{cc}
\delta_r   \\
\delta_i      \\
\end{array} } \right] = 
\left[ {\begin{array}{cc}
0  & \eta  \\
- \eta +4\mu g_1^2 & 0    \\
\end{array} } \right]
\left[ {\begin{array}{cc}
\delta_r   \\
\delta_i      \\
\end{array} } \right],
\end{equation}
where $\eta = \omega + 4\mu g_3^2+4\mu g_3$.  The
eigenvalues of this equation can be easily found to be,
 \begin{equation}
 \lambda = \pm \sqrt{\eta(4\mu g_1^2-\eta)}.
  \end{equation}
  Assuming $\eta>0$,  unstable solutions can only exist if $4\mu g_1^2>\eta$, meaning that $g_1^2>\omega/4\mu+g_3^2+g_3$ .
  This simple argument obviously shows the role of $\omega$ and $g_3$ in suppressing
  flavor instabilities discussed in the previous section, caused by the flavor-violating  $\nu$NSSI. 
   Note, however, 
  that this is the insight obtained ignoring the role of the $2\mu g_1(1+g_3)$ term, and  post our understanding regarding the roles of $g_1$, $g_3$, and $\omega$ 
  in the flavor instabilities. Otherwise  this simple argument has its own limitations, e.g., it fails  when $\eta<0$. 
  
  It should be kept in mind that the linear  stability analysis presented here reaches very
    different results  from the one in Ref.~\cite{Dighe:2017sur}, given
  the following observations: (i) The trace of $\rho$ is not removed in this study, as it should not be 
  (ii) The $ \mu g_1^2 \delta^*$ term in the linearised equation of motion is the source of instability, meaning
  that one should not drop it in favor of the other terms. Then, for example, we here observe that the 
  neutrino gas is always unstable as long as $g_1\gtrsim g_3$, 
  which is different from the results presented, e.g., in Fig.~1 of Ref.~\cite{Dighe:2017sur}.

\bibliographystyle{elsarticle-num}
\bibliography{NSI}

\begin{thebibliography}{10}
\expandafter\ifx\csname url\endcsname\relax
  \def\url#1{\texttt{#1}}\fi
\expandafter\ifx\csname urlprefix\endcsname\relax\def\urlprefix{URL }\fi
\expandafter\ifx\csname href\endcsname\relax
  \def\href#1#2{#2} \def\path#1{#1}\fi

\bibitem{Colgate:1966ax}
S.~A. Colgate, R.~H. White, {The Hydrodynamic Behavior of Supernovae
  Explosions}, Astrophys. J. 143 (1966) 626.
\newblock \href {https://doi.org/10.1086/148549} {\path{doi:10.1086/148549}}.

\bibitem{Bethe:1984ux}
H.~A. Bethe, J.~R. Wilson, {Revival of a stalled supernova shock by neutrino
  heating}, Astrophys. J. 295 (1985) 14--23.
\newblock \href {https://doi.org/10.1086/163343} {\path{doi:10.1086/163343}}.

\bibitem{Janka:2012wk}
H.-T. Janka, {Explosion Mechanisms of Core-Collapse Supernovae}, Ann. Rev.
  Nucl. Part. Sci. 62 (2012) 407--451.
\newblock \href {http://arxiv.org/abs/1206.2503} {\path{arXiv:1206.2503}},
  \href {https://doi.org/10.1146/annurev-nucl-102711-094901}
  {\path{doi:10.1146/annurev-nucl-102711-094901}}.

\bibitem{Burrows:2012ew}
A.~Burrows, {Colloquium: Perspectives on core-collapse supernova theory}, Rev.
  Mod. Phys. 85 (2013) 245.
\newblock \href {http://arxiv.org/abs/1210.4921} {\path{arXiv:1210.4921}},
  \href {https://doi.org/10.1103/RevModPhys.85.245}
  {\path{doi:10.1103/RevModPhys.85.245}}.

\bibitem{Qian:1996xt}
Y.~Z. Qian, S.~E. Woosley, {Nucleosynthesis in neutrino driven winds: 1. The
  Physical conditions}, Astrophys. J. 471 (1996) 331--351.
\newblock \href {http://arxiv.org/abs/astro-ph/9611094}
  {\path{arXiv:astro-ph/9611094}}, \href {https://doi.org/10.1086/177973}
  {\path{doi:10.1086/177973}}.

\bibitem{Pastor:2002we}
S.~Pastor, G.~Raffelt, Flavor oscillations in the supernova hot bubble region:
  Nonlinear effects of neutrino background, Phys. Rev. Lett. 89 (2002) 191101.
\newblock \href {http://arxiv.org/abs/astro-ph/0207281}
  {\path{arXiv:astro-ph/0207281}}.

\bibitem{duan:2006an}
H.~Duan, G.~M. Fuller, J.~Carlson, Y.-Z. Qian, {Simulation of Coherent
  Non-Linear Neutrino Flavor Transformation in the Supernova Environment. 1.
  Correlated Neutrino Trajectories}, Phys. Rev. D74 (2006) 105014.
\newblock \href {http://arxiv.org/abs/astro-ph/0606616}
  {\path{arXiv:astro-ph/0606616}}, \href
  {https://doi.org/10.1103/PhysRevD.74.105014}
  {\path{doi:10.1103/PhysRevD.74.105014}}.

\bibitem{duan:2006jv}
H.~Duan, G.~M. Fuller, J.~Carlson, Y.-Z. Qian, {Coherent Development of
  Neutrino Flavor in the Supernova Environment}, Phys. Rev. Lett. 97 (2006)
  241101.
\newblock \href {http://arxiv.org/abs/astro-ph/0608050}
  {\path{arXiv:astro-ph/0608050}}, \href
  {https://doi.org/10.1103/PhysRevLett.97.241101}
  {\path{doi:10.1103/PhysRevLett.97.241101}}.

\bibitem{duan:2010bg}
H.~Duan, G.~M. Fuller, Y.-Z. Qian, {Collective Neutrino Oscillations}, Ann.
  Rev. Nucl. Part. Sci. 60 (2010) 569.
\newblock \href {http://arxiv.org/abs/1001.2799} {\path{arXiv:1001.2799}}.

\bibitem{Chakraborty:2016yeg}
S.~Chakraborty, R.~Hansen, I.~Izaguirre, G.~Raffelt, {Collective neutrino
  flavor conversion: Recent developments}, Nucl. Phys. B908 (2016) 366--381.
\newblock \href {http://arxiv.org/abs/1602.02766} {\path{arXiv:1602.02766}},
  \href {https://doi.org/10.1016/j.nuclphysb.2016.02.012}
  {\path{doi:10.1016/j.nuclphysb.2016.02.012}}.

\bibitem{Bialynicka-Birula:1964ddi}
Z.~Bialynicka-Birula, {Do Neutrinos Interact between Themselves?}, Nuovo Cim.
  33 (1964) 1484--1487.
\newblock \href {https://doi.org/10.1007/BF02749481}
  {\path{doi:10.1007/BF02749481}}.

\bibitem{Bardin:1970wq}
D.~Y. Bardin, S.~M. Bilenky, B.~Pontecorvo, {On the nu - nu interaction}, Phys.
  Lett. B 32 (1970) 121--124.
\newblock \href {https://doi.org/10.1016/0370-2693(70)90602-7}
  {\path{doi:10.1016/0370-2693(70)90602-7}}.

\bibitem{Ge:2018uhz}
S.-F. Ge, S.~J. Parke, {Scalar Nonstandard Interactions in Neutrino
  Oscillation}, Phys. Rev. Lett. 122~(21) (2019) 211801.
\newblock \href {http://arxiv.org/abs/1812.08376} {\path{arXiv:1812.08376}},
  \href {https://doi.org/10.1103/PhysRevLett.122.211801}
  {\path{doi:10.1103/PhysRevLett.122.211801}}.

\bibitem{Yang:2018yvk}
Y.~Yang, J.~P. Kneller, {Neutrino flavor transformation in supernovae as a
  probe for nonstandard neutrino-scalar interactions}, Phys. Rev. D 97~(10)
  (2018) 103018.
\newblock \href {http://arxiv.org/abs/1803.04504} {\path{arXiv:1803.04504}},
  \href {https://doi.org/10.1103/PhysRevD.97.103018}
  {\path{doi:10.1103/PhysRevD.97.103018}}.

\bibitem{Berryman:2022hds}
J.~M. Berryman, et~al., {Neutrino Self-Interactions: A White Paper}, in: {2022
  Snowmass Summer Study}, 2022.
\newblock \href {http://arxiv.org/abs/2203.01955} {\path{arXiv:2203.01955}}.

\bibitem{Bilenky:1992xn}
M.~S. Bilenky, S.~M. Bilenky, A.~Santamaria, {Invisible width of the Z boson
  and 'secret' neutrino-neutrino interactions}, Phys. Lett. B 301 (1993)
  287--291.
\newblock \href {https://doi.org/10.1016/0370-2693(93)90703-K}
  {\path{doi:10.1016/0370-2693(93)90703-K}}.

\bibitem{Bilenky:1994ma}
M.~S. Bilenky, A.~Santamaria, {Bounding effective operators at the one loop
  level: The Case of four fermion neutrino interactions}, Phys. Lett. B 336
  (1994) 91--99.
\newblock \href {http://arxiv.org/abs/hep-ph/9405427}
  {\path{arXiv:hep-ph/9405427}}, \href
  {https://doi.org/10.1016/0370-2693(94)00961-9}
  {\path{doi:10.1016/0370-2693(94)00961-9}}.

\bibitem{Brdar:2020nbj}
V.~Brdar, M.~Lindner, S.~Vogl, X.-J. Xu, {Revisiting neutrino self-interaction
  constraints from $Z$ and $\tau$ decays}, Phys. Rev. D 101~(11) (2020) 115001.
\newblock \href {http://arxiv.org/abs/2003.05339} {\path{arXiv:2003.05339}},
  \href {https://doi.org/10.1103/PhysRevD.101.115001}
  {\path{doi:10.1103/PhysRevD.101.115001}}.

\bibitem{Ahlgren:2013wba}
B.~Ahlgren, T.~Ohlsson, S.~Zhou, {Comment on \textquotedblleft{}Is Dark Matter
  with Long-Range Interactions a Solution to All Small-Scale Problems of
  \ensuremath{\Lambda} Cold Dark Matter Cosmology?\textquotedblright{}}, Phys.
  Rev. Lett. 111~(19) (2013) 199001.
\newblock \href {http://arxiv.org/abs/1309.0991} {\path{arXiv:1309.0991}},
  \href {https://doi.org/10.1103/PhysRevLett.111.199001}
  {\path{doi:10.1103/PhysRevLett.111.199001}}.

\bibitem{Archidiacono:2013dua}
M.~Archidiacono, S.~Hannestad, {Updated constraints on non-standard neutrino
  interactions from Planck}, JCAP 07 (2014) 046.
\newblock \href {http://arxiv.org/abs/1311.3873} {\path{arXiv:1311.3873}},
  \href {https://doi.org/10.1088/1475-7516/2014/07/046}
  {\path{doi:10.1088/1475-7516/2014/07/046}}.

\bibitem{Cyr-Racine:2013jua}
F.-Y. Cyr-Racine, K.~Sigurdson, {Limits on Neutrino-Neutrino Scattering in the
  Early Universe}, Phys. Rev. D 90~(12) (2014) 123533.
\newblock \href {http://arxiv.org/abs/1306.1536} {\path{arXiv:1306.1536}},
  \href {https://doi.org/10.1103/PhysRevD.90.123533}
  {\path{doi:10.1103/PhysRevD.90.123533}}.

\bibitem{Esteban:2021tub}
I.~Esteban, S.~Pandey, V.~Brdar, J.~F. Beacom, {Probing secret interactions of
  astrophysical neutrinos in the high-statistics era}, Phys. Rev. D 104~(12)
  (2021) 123014.
\newblock \href {http://arxiv.org/abs/2107.13568} {\path{arXiv:2107.13568}},
  \href {https://doi.org/10.1103/PhysRevD.104.123014}
  {\path{doi:10.1103/PhysRevD.104.123014}}.

\bibitem{Ioka:2014kca}
K.~Ioka, K.~Murase, {IceCube PeV\textendash{}EeV neutrinos and secret
  interactions of neutrinos}, PTEP 2014~(6) (2014) 061E01.
\newblock \href {http://arxiv.org/abs/1404.2279} {\path{arXiv:1404.2279}},
  \href {https://doi.org/10.1093/ptep/ptu090} {\path{doi:10.1093/ptep/ptu090}}.

\bibitem{Ng:2014pca}
K.~C.~Y. Ng, J.~F. Beacom, {Cosmic neutrino cascades from secret neutrino
  interactions}, Phys. Rev. D 90~(6) (2014) 065035, [Erratum: Phys.Rev.D 90,
  089904 (2014)].
\newblock \href {http://arxiv.org/abs/1404.2288} {\path{arXiv:1404.2288}},
  \href {https://doi.org/10.1103/PhysRevD.90.065035}
  {\path{doi:10.1103/PhysRevD.90.065035}}.

\bibitem{Ibe:2014pja}
M.~Ibe, K.~Kaneta, {Cosmic neutrino background absorption line in the neutrino
  spectrum at IceCube}, Phys. Rev. D 90~(5) (2014) 053011.
\newblock \href {http://arxiv.org/abs/1407.2848} {\path{arXiv:1407.2848}},
  \href {https://doi.org/10.1103/PhysRevD.90.053011}
  {\path{doi:10.1103/PhysRevD.90.053011}}.

\bibitem{2014arXiv1408.3799B}
K.~{Blum}, A.~{Hook}, K.~{Murase}, {High energy neutrino telescopes as a probe
  of the neutrino mass mechanism}, arXiv e-prints (2014) arXiv:1408.3799\href
  {http://arxiv.org/abs/1408.3799} {\path{arXiv:1408.3799}}.

\bibitem{DiFranzo:2015qea}
A.~DiFranzo, D.~Hooper, {Searching for MeV-Scale Gauge Bosons with IceCube},
  Phys. Rev. D 92~(9) (2015) 095007.
\newblock \href {http://arxiv.org/abs/1507.03015} {\path{arXiv:1507.03015}},
  \href {https://doi.org/10.1103/PhysRevD.92.095007}
  {\path{doi:10.1103/PhysRevD.92.095007}}.

\bibitem{Shoemaker:2015qul}
I.~M. Shoemaker, K.~Murase, {Probing BSM Neutrino Physics with Flavor and
  Spectral Distortions: Prospects for Future High-Energy Neutrino Telescopes},
  Phys. Rev. D 93~(8) (2016) 085004.
\newblock \href {http://arxiv.org/abs/1512.07228} {\path{arXiv:1512.07228}},
  \href {https://doi.org/10.1103/PhysRevD.93.085004}
  {\path{doi:10.1103/PhysRevD.93.085004}}.

\bibitem{Kolb:1987qy}
E.~W. Kolb, M.~S. Turner, {Supernova SN 1987a and the Secret Interactions of
  Neutrinos}, Phys. Rev. D 36 (1987) 2895.
\newblock \href {https://doi.org/10.1103/PhysRevD.36.2895}
  {\path{doi:10.1103/PhysRevD.36.2895}}.

\bibitem{Shalgar:2019rqe}
S.~Shalgar, I.~Tamborra, M.~Bustamante, {Core-collapse supernovae stymie secret
  neutrino interactions}, Phys. Rev. D 103~(12) (2021) 123008.
\newblock \href {http://arxiv.org/abs/1912.09115} {\path{arXiv:1912.09115}},
  \href {https://doi.org/10.1103/PhysRevD.103.123008}
  {\path{doi:10.1103/PhysRevD.103.123008}}.

\bibitem{Chang:2022aas}
P.-W. Chang, I.~Esteban, J.~F. Beacom, T.~A. Thompson, C.~M. Hirata, {Towards
  Powerful Probes of Neutrino Self-Interactions in Supernovae} (6 2022).
\newblock \href {http://arxiv.org/abs/2206.12426} {\path{arXiv:2206.12426}}.

\bibitem{Sigl:1992fn}
G.~Sigl, G.~Raffelt, {General kinetic description of relativistic mixed
  neutrinos}, Nucl. Phys. B406 (1993) 423--451.
\newblock \href {https://doi.org/10.1016/0550-3213(93)90175-O}
  {\path{doi:10.1016/0550-3213(93)90175-O}}.

\bibitem{Wolfenstein:1977ue}
L.~Wolfenstein, {Neutrino Oscillations in Matter}, Phys. Rev. D17 (1978)
  2369--2374, [,294(1977)].
\newblock \href {https://doi.org/10.1103/PhysRevD.17.2369}
  {\path{doi:10.1103/PhysRevD.17.2369}}.

\bibitem{Mikheev:1986gs}
S.~P. Mikheyev, A.~{\relax Yu}. Smirnov, {Resonance Amplification of
  Oscillations in Matter and Spectroscopy of Solar Neutrinos}, Sov. J. Nucl.
  Phys. 42 (1985) 913--917, [,305(1986)].

\bibitem{sigl1993general}
G.~Sigl, G.~Raffelt, General kinetic description of relativistic mixed
  neutrinos, Nuclear Physics B 406~(1-2) (1993) 423--451.

\bibitem{Blennow:2008er}
M.~Blennow, A.~Mirizzi, P.~D. Serpico, {Nonstandard neutrino-neutrino
  refractive effects in dense neutrino gases}, Phys. Rev. D 78 (2008) 113004.
\newblock \href {http://arxiv.org/abs/0810.2297} {\path{arXiv:0810.2297}},
  \href {https://doi.org/10.1103/PhysRevD.78.113004}
  {\path{doi:10.1103/PhysRevD.78.113004}}.

\bibitem{Fuller:1987aa}
G.~M. Fuller, R.~W. Mayle, J.~R. Wilson, D.~N. Schramm, Resonant neutrino
  oscillations and stellar collapse, Astrophys. J. 322 (1987) 795.

\bibitem{Pantaleone:1992xh}
J.~T. Pantaleone, {Dirac neutrinos in dense matter}, Phys. Rev. D46 (1992)
  510--523.
\newblock \href {https://doi.org/10.1103/PhysRevD.46.510}
  {\path{doi:10.1103/PhysRevD.46.510}}.

\bibitem{Notzold:1988kx}
D.~N\"{o}tzold, G.~Raffelt, Neutrono dispersion at finite temperature and
  density, Nucl. Phys. B307 (1988) 924.

\bibitem{Das:2017iuj}
A.~Das, A.~Dighe, M.~Sen, {New effects of non-standard self-interactions of
  neutrinos in a supernova}, JCAP 05 (2017) 051.
\newblock \href {http://arxiv.org/abs/1705.00468} {\path{arXiv:1705.00468}},
  \href {https://doi.org/10.1088/1475-7516/2017/05/051}
  {\path{doi:10.1088/1475-7516/2017/05/051}}.

\bibitem{Dighe:2017sur}
A.~Dighe, M.~Sen, {Nonstandard neutrino self-interactions in a supernova and
  fast flavor conversions}, Phys. Rev. D 97~(4) (2018) 043011.
\newblock \href {http://arxiv.org/abs/1709.06858} {\path{arXiv:1709.06858}},
  \href {https://doi.org/10.1103/PhysRevD.97.043011}
  {\path{doi:10.1103/PhysRevD.97.043011}}.

\bibitem{Abbar:2018beu}
S.~Abbar, M.~C. Volpe, {On Fast Neutrino Flavor Conversion Modes in the
  Nonlinear Regime}, Phys. Lett. B790 (2019) 545--550.
\newblock \href {http://arxiv.org/abs/1811.04215} {\path{arXiv:1811.04215}},
  \href {https://doi.org/10.1016/j.physletb.2019.02.002}
  {\path{doi:10.1016/j.physletb.2019.02.002}}.

\bibitem{Abbar:2015mca}
S.~Abbar, H.~Duan, S.~Shalgar, {Flavor instabilities in the multiangle neutrino
  line model}, Phys. Rev. D 92~(6) (2015) 065019.
\newblock \href {http://arxiv.org/abs/1507.08992} {\path{arXiv:1507.08992}},
  \href {https://doi.org/10.1103/PhysRevD.92.065019}
  {\path{doi:10.1103/PhysRevD.92.065019}}.

\bibitem{Martin:2019kgi}
J.~D. Martin, S.~Abbar, H.~Duan, {Nonlinear flavor development of a
  two-dimensional neutrino gas}, Phys. Rev. D100~(2) (2019) 023016.
\newblock \href {http://arxiv.org/abs/1904.08877} {\path{arXiv:1904.08877}},
  \href {https://doi.org/10.1103/PhysRevD.100.023016}
  {\path{doi:10.1103/PhysRevD.100.023016}}.

\bibitem{Abbar:2020ror}
S.~Abbar, {Turbulence Fingerprint on Collective Oscillations of Supernova
  Neutrinos} (7 2020).
\newblock \href {http://arxiv.org/abs/2007.13655} {\path{arXiv:2007.13655}}.

\bibitem{Duan:2007sh}
H.~Duan, G.~M. Fuller, J.~Carlson, Y.-Z. Qian, {Flavor Evolution of the
  Neutronization Neutrino Burst from an O-Ne-Mg Core-Collapse Supernova}, Phys.
  Rev. Lett. 100 (2008) 021101.
\newblock \href {http://arxiv.org/abs/0710.1271} {\path{arXiv:0710.1271}},
  \href {https://doi.org/10.1103/PhysRevLett.100.021101}
  {\path{doi:10.1103/PhysRevLett.100.021101}}.

\bibitem{Dasgupta:2008cd}
B.~Dasgupta, A.~Dighe, A.~Mirizzi, G.~G. Raffelt, {Spectral split in prompt
  supernova neutrino burst: Analytic three-flavor treatment}, Phys. Rev. D 77
  (2008) 113007.
\newblock \href {http://arxiv.org/abs/0801.1660} {\path{arXiv:0801.1660}},
  \href {https://doi.org/10.1103/PhysRevD.77.113007}
  {\path{doi:10.1103/PhysRevD.77.113007}}.

\bibitem{Proceedings:2019qno}
{Neutrino Non-Standard Interactions: A Status Report}, Vol.~2.
\newblock \href {http://arxiv.org/abs/1907.00991} {\path{arXiv:1907.00991}},
  \href {https://doi.org/10.21468/SciPostPhysProc.2.001}
  {\path{doi:10.21468/SciPostPhysProc.2.001}}.

\bibitem{Biggio:2009nt}
C.~Biggio, M.~Blennow, E.~Fernandez-Martinez, {General bounds on non-standard
  neutrino interactions}, JHEP 08 (2009) 090.
\newblock \href {http://arxiv.org/abs/0907.0097} {\path{arXiv:0907.0097}},
  \href {https://doi.org/10.1088/1126-6708/2009/08/090}
  {\path{doi:10.1088/1126-6708/2009/08/090}}.

\bibitem{Stapleford:2016jgz}
C.~J. Stapleford, D.~J. V\"a\"an\"anen, J.~P. Kneller, G.~C. McLaughlin, B.~T.
  Shapiro, {Nonstandard Neutrino Interactions in Supernovae}, Phys. Rev. D
  94~(9) (2016) 093007.
\newblock \href {http://arxiv.org/abs/1605.04903} {\path{arXiv:1605.04903}},
  \href {https://doi.org/10.1103/PhysRevD.94.093007}
  {\path{doi:10.1103/PhysRevD.94.093007}}.

\bibitem{Chatelain:2017yxx}
A.~Chatelain, M.~C. Volpe, {Neutrino propagation in binary neutron star mergers
  in presence of nonstandard interactions}, Phys. Rev. D 97~(2) (2018) 023014.
\newblock \href {http://arxiv.org/abs/1710.11518} {\path{arXiv:1710.11518}},
  \href {https://doi.org/10.1103/PhysRevD.97.023014}
  {\path{doi:10.1103/PhysRevD.97.023014}}.

\bibitem{Babu:2019iml}
K.~S. Babu, G.~Chauhan, P.~S. Bhupal~Dev, {Neutrino nonstandard interactions
  via light scalars in the Earth, Sun, supernovae, and the early Universe},
  Phys. Rev. D 101~(9) (2020) 095029.
\newblock \href {http://arxiv.org/abs/1912.13488} {\path{arXiv:1912.13488}},
  \href {https://doi.org/10.1103/PhysRevD.101.095029}
  {\path{doi:10.1103/PhysRevD.101.095029}}.

\bibitem{Fogli:2002xj}
G.~L. Fogli, E.~Lisi, A.~Mirizzi, D.~Montanino, {Revisiting nonstandard
  interaction effects on supernova neutrino flavor oscillations}, Phys. Rev. D
  66 (2002) 013009.
\newblock \href {http://arxiv.org/abs/hep-ph/0202269}
  {\path{arXiv:hep-ph/0202269}}, \href
  {https://doi.org/10.1103/PhysRevD.66.013009}
  {\path{doi:10.1103/PhysRevD.66.013009}}.

\end{thebibliography}

\clearpage

\end{document}